\documentclass[twocolumn,amsmath,amssymb,prb,floatfix]{revtex4}

\usepackage{graphicx}   
\usepackage{dcolumn}    
\usepackage{bm}         
\usepackage{rotating}
\usepackage{enumerate}

\newcommand{\cpl}{Chem.\ Phys.\ Lett.\ }

\renewcommand{\jcp}{J.\ Chem.\ Phys.\ }

\newcommand{\jpc}{J.\ Phys.\ Chem.\ }
\newcommand{\jpca}{J.\ Phys.\ Chem.\ A }
\newcommand{\jpcc}{J.\ Phys.\ Chem.\ C }

\renewcommand{\prl}{Phys.\ Rev.\ Lett.\ }

\newcommand{\jpcl}{J.\ Phys.\ Chem.\ Lett.\ }

\newcommand{\pnas}{Proc.\ Natl.\ Acad.\ Sci.\ }
\newcommand{\eqn}[1]{Eq.~(\ref{#1})}
\newcommand{\Eqn}[1]{Equation~(\ref{#1})}
\newcommand{\eqnn}[2]{Eqs.~(\ref{#1}) and (\ref{#2})}

\newcommand{\bra}[1]{\big< \, #1 \, \big|}
\newcommand{\ket}[1]{\big| \, #1 \, \big>}
\newcommand{\expect}[1]{\big< \, #1 \, \big>}

\newcommand{\w}{\widetilde}
\newcommand{\no}{\nonumber}
\newcommand{\shortt}{$t \to 0_+$\ }
\newcommand{\shortta}{$t \to 0_+$}
\newcommand{\longt}{$t \to \infty$\ }
\newcommand{\longta}{$t \to \infty$}
\newcommand{\nty}{$N \to \infty$\ }
\newcommand{\ntya}{$N \to \infty$}

\newcommand{\ppi}{{\boldsymbol \pi}}
\newcommand{\ssi}{{\boldsymbol \sigma}}
\newcommand{\ov}{\overline}

\begin{document} 

\title{Derivation of a true (${t\rightarrow 0_+}$) quantum transition-state theory. \\II. Recovery of the exact quantum rate in the absence of recrossing} 
\author{Stuart C.~Althorpe\footnote{Corresponding author: sca10@cam.ac.uk} and Timothy J.~H.~Hele}
\affiliation{Department of Chemistry, University of Cambridge, Lensfield Road, Cambridge, CB2 1EW, UK.}
\date{\today}

\begin{abstract} 
In Part I [J.\ Chem.\ Phys.\ {\bf 138}, 084108 (2013)] we derived a quantum transition-state theory by taking the \shortt limit of a new form of quantum flux-side time-correlation function containing a ring-polymer dividing surface. This \shortt limit appears to be unique in giving positive-definite Boltzmann statistics,  and is {\em identical} to ring-polymer molecular dynamics (RPMD) TST. Here, we show that quantum TST (i.e.\ RPMD-TST) is exact if there is no recrossing (by the real-time quantum dynamics) of the ring-polymer dividing surface, nor of any surface orthogonal to it in the space describing fluctuations in the polymer-bead positions along the reaction coordinate. In practice, this means that RPMD-TST gives a good approximation to the exact quantum rate for direct reactions, provided the temperature is not too far below the cross-over to deep tunnelling.
We derive these results by comparing the
\longt limit of the ring-polymer flux-side time-correlation function with that of a hybrid flux-side time-correlation function  (containing a ring-polymer flux operator and a Miller-Schwarz-Tromp side function), and by representing the resulting ring-polymer momentum integrals as hypercubes. Together with Part I, the results of this article validate a large number of RPMD calculations of reaction rates.
\emph{Copyright (2013) American Institute of Physics. This article may be downloaded for personal use only. Any other use requires prior permission of the author and the American Institute of Physics. The following article appeared in The Journal of Chemical Physics, \textbf{139} (2013) 084115, and may be found at http://link.aip.org/link/?JCP/139/084115/1}
\end{abstract}

\maketitle

\section{Introduction}

In Part I, \cite{part1} we derived a quantum generalization of classical transition-state theory (TST), which corresponds to the \shortt limit of a new form of quantum flux-side time-correlation function. This function uses a ring-polymer \cite{chandler} dividing surface, which is invariant under cyclic permutation of the polymer beads, and thus becomes invariant to imaginary-time translation in the infinite-bead limit.  The resulting quantum TST appears to be unique,  in the sense that the \shortt limit of any other known form of flux-side time-correlation function \cite{part1,MST,mill,pollack} gives either incorrect quantum statistics, or zero.  Remarkably, this quantum TST is {\em identical} to ring-polymer molecular dynamics (RPMD) TST, \cite{jor} and thus validates a large number of recent RPMD rate calculations,\cite{rates,refined,azzouz,bimolec,ch4,mustuff,anrev,yury,tommy1,tommy2,tommy3,stecher,guo} as well as the earlier-developed `quantum TST method'\cite{gillan1,gillan2,centroid1,centroid2,schwieters,ides} (which is RPMD-TST in the special case of a centroid dividing surface,\cite{jor}  and which, to avoid confusion,
we will refer to here as `centroid-TST'\cite{cent}).

 There are a variety of other  methods for estimating the quantum rate based on short-time\cite{pollack,scivr1,scivr2,scivr3,QI1,QI2} or
 semiclassical\cite{billact,billhandy,stanton,bill,cole,benders,jonss,spanish,kastner1,kastner2,equiv} dynamics. What is different about quantum TST is that it corresponds to the instantaneous $t\rightarrow 0_+$ quantum flux through a dividing surface.
Classical TST corresponds to the analogous \shortt classical flux, which is well known to give the exact (classical) rate if there is no recrossing of the dividing surface;\cite{green,daan} in practice, there is always some such recrossing, and thus classical TST gives a {\em good approximation} to the exact (classical) rate for systems in 
which the amount of recrossing is small, namely direct reactions. The purpose of this article is to derive the analogous result for quantum TST  (i.e.\ RPMD-TST), to show
that it gives the exact quantum rate if there is no recrossing (by the exact quantum dynamics\cite{nospring}), and thus that it gives a good 
approximation to the exact quantum rate for direct reactions.

To clarify the work ahead, we summarize two important differences between classical and quantum TST. First, classical TST gives a strict upper bound to the corresponding exact rate, but quantum TST does not, since real-time coherences may increase the quantum flux upon recrossing.\cite{part1} Quantum TST breaks down if such coherences are large; one then has no choice but to attempt to model the real-time quantum dynamics. However, in many systems (especially in the condensed phase), real-time quantum coherence has a negligible effect on the rate. In such systems, quantum TST gives a {\em good approximation to} an upper bound to the exact quantum rate. This becomes a strict upper bound only in the high-temperature limit, where classical TST is recovered as a special limiting case.

Second, when discussing recrossing in classical TST,  one has only to consider whether trajectories initiated on the dividing surface recross that surface. In quantum TST, the time-evolution operator is applied to a series of $N$ initial positions, corresponding to the positions of the polymer beads. A consequence of this, as we discuss below, is that one needs to consider, not just recrossing (by the exact quantum dynamics) of the ring-polymer dividing surface,
 but also of surfaces orthogonal to it in the ($N\!-\!1$)-dimensional space  describing fluctuations in the polymer-bead positions along the reaction coordinate.
A major task of this article will be to show that the recrossing of these surfaces (by the exact quantum dynamics) causes the long-time limit of the ring-polymer flux-side time-correlation function to differ from the exact quantum rate. It then follows that the RPMD-TST rate is equal to the exact quantum rate if there is neither recrossing of the ring-polymer dividing surface, nor of any  of these $N\!-\!1$ orthogonal surfaces.

We will use quantum scattering theory to derive these results, although we emphasise that they apply also in condensed phases (where RPMD has proved particularly groundbreaking \cite{azzouz,anrev,yury,tommy1,tommy2,tommy3}). The scattering theory is employed merely as a derivational tool, exploiting the property that the flux-side plateau in a scattering system extends to infinite time, which makes derivation of the rate straightforward. The results thus derived can be applied in the condensed phase, subject to the usual caveat of there being a separation in timescales between barrier-crossing and equilibration. \cite{isomer,linres} We have relegated most of the scattering theory to Appendices, in the hope that the outline of the derivation can be followed in the main body of the text.

The article is structured as follows: After summarizing the main findings of Part I in Sec.~II, we introduce in Sec.~III a hybrid flux-side time-correlation function, which correlates flux through the ring-polymer dividing surface with the Miller-Schwarz-Tromp\cite{MST} side function, and which gives the exact quantum rate in the  limit $t\rightarrow\infty$. We describe the $N$-dimensional integral over momenta obtained in this limit by an $N$-dimensional hypercube, and note that the $t\rightarrow\infty$ limits of the ring-polymer and hybrid flux-side time-correlation functions cut out different volumes from the hypercube, thus explaining why the former does not in general give the exact quantum rate. In Sec.~IV we show that the only parts of the integrand that cause this difference are
a series of Dirac $\delta$-function spikes running through the hypercube. In Sec.~V we show that these spikes  disappear if there is no recrossing (by the exact quantum dynamics\cite{nospring}) in the ($N\!-\!1$)-dimensional space orthogonal to the dividing surface (mentioned above). It then follows that the RPMD-TST rate is equal to the exact rate if there  is also no recrossing of the dividing surface itself. In Sec.~VI we explain how these results (which were derived in one dimension) generalize  to multi-dimensions. Section VII concludes the article.

\label{sec:intro}
\section{Summary of Part I}

Here we summarize the main results of Part I. To simplify the algebra, we focus on a one-dimensional scattering system with hamiltonian ${\hat H}$, potential $V(x)$ and mass $m$. However, the results generalize immediately  
 to multi-dimensional systems (see Sec.~VI) and to the condensed phase (see comments in the Introduction). 
 
The ring-polymer flux-side time-correlation function, introduced in Part I, is 
\begin{align}
C_{\rm fs}^{[N]}(t)=&\int\! d{\bf q}\, \int\! d{\bf z}\,\int\! d{\bf \Delta}\,{\cal \hat F}[f({\bf q})]h[f({\bf z})]\nonumber\\ 
&\times\prod_{i=1}^{N}\expect{q_{i-1}-\Delta_{i-1}/2|e^{-\beta_N{\hat H}}|q_i+\Delta_i/2}
\nonumber\\ 
&\quad\times \expect{q_i+\Delta_i/2|e^{i{\hat H}t/\hbar}|z_i} \nonumber \\
& \quad \times \expect{z_i|e^{-i{\hat H}t/\hbar}|q_i-\Delta_i/2}
\label{utter} 
\end{align}
where $N$ is the number of polymer beads, $\beta_N=\beta/N$, with $\beta=1/k_{\rm B}T$, and 
${\bf q}\equiv\{q_1,\dots,q_N\}$, with ${\bf z}$ and $\Delta$ similarly defined. The function $f({\bf q})$ is the ring-polymer dividing surface, which is invariant under cyclic permutations of the polymer beads (i.e.\ of the individual $q_i$), and thus becomes invariant to imaginary-time translation in the limit $N\rightarrow\infty$. The operator ${\cal \hat F}[f({\bf q})]$ gives the flux perpendicular to $f({\bf q})$, and is given by
\begin{align}
{\cal \hat F}[f({\bf q})] &= {1\over 2m}\sum_{i=1}^N\left\{{\hat p_i}{\partial f({\bf q})\over\partial q_i}  \delta[f({\bf q})]+  \delta[f({\bf q})]{\partial f({\bf q})\over\partial q_i}{\hat p_i}\right\}
\label{eq:superflux} 
\end{align}
Note that we employ here a convention introduced in Part I,  that the first term inside the curly brackets is inserted between $e^{-\beta_N \hat H}\ket{q_i+\Delta_i/2}$  and $\bra{q_i+\Delta_i/2}e^{i{\hat H}t/\hbar}$ in \eqn{utter},
and the second term between $e^{-i{\hat H}t/\hbar}\ket{q_i+\Delta_i/2}$ and $\bra{q_i+\Delta_i/2}e^{-\beta_N\hat H}$. This is done to emphasise the form of $C_{\rm fs}^{[N]}(t)$; [\eqn{utter} is written out in full in Part I].  

We can regard $C_{\rm fs}^{[N]}(t)$ as a generalized Kubo-transformed time-correlation function, since it
correlates an operator (in this case ${\cal \hat F}[f({\bf q})]$) on the (imaginary-time) Feynman paths at $t=0$ with another operator (in this case $h[f({\bf z})])$ at some later time $t$, and would reduce to a standard Kubo-transformed function if these operators were replaced by linear functions of position or momentum operators. The advantage of $C_{\rm fs}^{[N]}(t)$ is that it allows both the flux and the side dividing surface to be made the {\em same} function of ring-polymer space (i.e. $f$), which is what makes $C_{\rm fs}^{[N]}(t)$ non-zero in the limit $t\rightarrow\infty$. One can show\cite{part1} that the invariance of $f({\bf q})$ to imaginary time-translation in the limit $N\rightarrow\infty$ ensures that $C_{\rm fs}^{[N]}(t)$ is positive-definite in the 
limits ${t\rightarrow 0_+}$ and ${N\rightarrow\infty}$. This allows us to define the quantum TST rate
  \begin{align}
  k_{Q}^\ddag(\beta)Q_{\rm r}(\beta)=\lim_{t\rightarrow 0_+}\lim_{N\rightarrow\infty}C_{\rm fs}^{[N]}(t)
 \end{align} 
where
\begin{align}
k_{Q}^\ddag(\beta)Q_{\rm r}(\beta)=&\lim_{N\rightarrow\infty}{1\over (2\pi\hbar)^N}\int\! d{\bf q}\, \int\! dP_0\, \delta[f({\bf q})] \nonumber \\
&\times \sqrt{B_N({\bf q})}\frac{P_0}{m}h\!\left(P_0\right)\sqrt{2\pi\beta_N\hbar^2\over m} \nonumber\\ 
& \times e^{-P_0^2\beta_N/2m}\prod_{i=1}^{N}\expect{q_{i-1}|e^{-\beta_N{\hat H}}|q_i}\label{bog}
\end{align}
Comparison with refs.~\onlinecite{jor,rates,refined} shows that $k_{Q}^\ddag(\beta)$ is {\em identical} to the RPMD-TST rate. The terms `quantum TST' and `RPMD-TST' are therefore equivalent (and will be used interchangeably throughout the article).

For quantum TST to be applicable, one must be able to assume that real-time coherences have only a small effect on the rate. It then follows that (a good approximation to) the optimal dividing surface $f({\bf q})$ is the one that maximises the free energy of the ring-polymer ensemble. If the reaction barrier is reasonably symmetric,\cite{vlow} or if it is asymmetric but the temperature is too hot for deep tunnelling, then a good choice of dividing surface is
\begin{align}
f({\bf q}) = {\overline q}_0-q^\ddag
\end{align} 
where 
\begin{align}
{\overline q}_0 = {1\over N}\sum_{i=1}^Nq_i
\end{align}
is the centroid.
(This special case of RPMD-TST was introduced earlier\cite{gillan1,gillan2,centroid1,centroid2,schwieters} and referred to as `quantum TST'; to avoid confusion we refer to it here as `centroid-TST' \cite{cent}) If the barrier is asymmetric, and the temperature is below the cross-over to deep tunnelling, then a more complicated dividing surface should be used which allows the polymer to stretch.\cite{jor}  As mentioned above, $f({\bf q})$ must be invariant under cyclic permutation of the beads so that it becomes invariant to imaginary time-translation in  the limit $N\rightarrow\infty$, and thus gives positive-definite quantum statistics. 

It is assumed above, and was stated without proof in Part I, that the RPMD-TST rate gives the exact quantum rate in the absence of recrossing, and is thus a good approximation to the exact rate if the amount of recrossing is small. The remainder of this article is devoted to deriving this result.

\section{Long-time limits}
\subsection{Hybrid flux-side time-correlation function}
To analyze the $t\rightarrow\infty$ limit of $C_{\rm fs}^{[N]}(t)$, we will find it convenient to consider  the $t\rightarrow\infty$ limit of the closely related {\em hybrid} flux-side time-correlation function:
\begin{align}
{\overline C}_{\rm fs}^{[N]}(t)= & \int\! d{\bf q}\, \int\! d{\bf z}\,\int\! d{\bf \Delta}\,{\cal \hat F}[f({\bf q})]h(z_1-q^\ddag)\nonumber\\ 
& \times \prod_{i=1}^{N}\expect{q_{i-1}-\Delta_{i-1}/2|e^{-\beta_N{\hat H}}|q_i+\Delta_i/2}
\nonumber\\ 
& \quad\times \expect{q_i+\Delta_i/2|e^{i{\hat H}t/\hbar}|z_i} \nonumber \\
& \quad\times \expect{z_i|e^{-i{\hat H}t/\hbar}|q_i-\Delta_i/2}
\label{gruel} 
\end{align}
 Note that we could equivalently have inserted any one of the other $z_i$ into the side-function, and also that we could simplify this expression by
collapsing the identities $\int\! dz_i\, e^{i{\hat H}t/\hbar}\ket{z_i}
\bra{z_i}e^{-i{\hat H}t/\hbar}$, $ i\ne 1$ [but we have not done so in order to emphasise the relation with ${ C}_{\rm fs}^{[N]}(t)$].

The function ${\overline C}_{\rm fs}^{[N]}(t)$ does not give a quantum TST, except in the special case that $N=1$ and $f({\bf q})=q_1$. In this case,
 ${\overline C}_{\rm fs}^{[N]}(t)$ is identical to $C_{\rm fs}^{[1]}(t)$, whose \shortt\ limit was shown in Part I to be identical to
 the quantum TST introduced on heuristic grounds by Wigner in 1932.\cite{wiggy} For $N>1$, 
 the flux and side dividing surfaces in ${\overline C}_{\rm fs}^{[N]}(t)$ are different functions of ring-polymer space, with the result that ${\overline C}_{\rm fs}^{[N]}(t)$ tends smoothly to zero in the limit \shortta. \cite{part1}

By taking the $t\to\infty$ limit of the equivalent {\em side-flux}
 time-correlation function ${\overline C}_{\rm sf}^{[N]}(t)$, we show in Appendix A that
\begin{align}
k_{Q}(\beta)Q_{\rm r}(\beta)=\lim_{t\rightarrow\infty}{\overline C}_{\rm fs}^{[N]}(t)
\label{thicky}
\end{align}
where $k_Q(\beta)$ is the exact quantum rate, and this expression holds for all $N\ge 1$. For $N=1$, we have thus proved that the flux-side time-correlation function
that gives  the Wigner form of quantum TST (see above) also gives the exact rate in the limit $t\to\infty$. \cite{23}  For $N>1$, which is  our main concern here, ${\overline C}_{\rm fs}^{[N]}(t)$ has the same limits as the Miller-Schwarz-Tromp\cite{MST} flux-side time-correlation function, tending smoothly to zero as \shortta, and  giving the exact quantum rate as $t\to \infty$. We can also evaluate the $t\to \infty$ limit of ${\overline C}_{\rm fs}^{[N]}(t)$ directly [i.e.\ not via ${\overline C}_{\rm sf}^{[N]}(t)$]. We apply
first the relation
\begin{align}
\lim_{t\to\infty} &\int_{-\infty}^\infty \! dz\, \expect{x|e^{i\hat K t/\hbar}|z}h(z-q^\ddag)\expect{z|e^{-i\hat K t/\hbar}|y}=\nonumber\\
&\int_{-\infty}^\infty \! dp\,\expect{\!x|p\!}h(p)\expect{\!p|y\!}\label{dragons}
\end{align}
where ${\hat K}$ is the kinetic energy operator and $\expect{x|p} = (2\pi\hbar)^{-1/2}\exp{(ipx)}$; 
this converts \eqn{gruel} into a form that involves applications of the M{\o }ller operator\cite{taylor}
\begin{align}
\hat\Omega_-\equiv\lim_{t\to\infty}e^{i\hat H t/\hbar}e^{-i\hat K t/\hbar}
\end{align}
onto momentum states $\ket{p_i}$. We then use the relation
\begin{align}
\hat\Omega_-\ket{p}= \ket{\phi^-_{p}}
\end{align}
where $\ket{\phi^-_{p}}$ is the (reactive) scattering wave function with outgoing boundary conditions, \cite{bc} to obtain
\begin{align}
\lim_{t\rightarrow \infty} {\overline C}_{\rm fs}^{[N]}(t)=\int\! d{\bf p}\,A_N({\bf p})h(p_1)
\label{eq:longtana} 
\end{align}
 with
\begin{align}
A_N({\bf p})=\int\! d{\bf q}\, &\int\! d{\bf \Delta}\,{\cal \hat F}[f({\bf q})]\nonumber\\ 
\times&\prod_{i=1}^{N}\expect{q_{i-1}-\Delta_{i-1}/2|e^{-\beta_N{\hat H}}|q_i+\Delta_i/2}
\nonumber\\ 
\quad\times &\expect{q_i+\Delta_i/2|\phi^-_{p_i}}
\expect{\phi^-_{p_i}|q_i-\Delta_i/2}\label{pete}
 \end{align}
 
 \subsection{Representation of the ring-polymer momentum integral}  
 To analyze the properties of \eqn{eq:longtana} (and of \eqn{boodles} given below),  we will find it helpful to 
 represent the space occupied by the intregrand as an $N$-dimensional hypercube,\cite{hyper} whose edges are the axes
 $-p_{\rm max}<p_i<p_{\rm max}$, $i=1\dots N$, in the limit $p_{\rm max}\rightarrow\infty$. We assume no familiarity with the geometry of hypercubes, and in fact use this terminology mainly to indicate that once a property of $A_N({\bf p})$ has been derived for $N=3$ (where the hypercube is simply a cube and thus easily visualised as in Fig.~1) it generalizes straightforwardly to higher $N$. 
 
 \newlength{\figwidths}
\setlength{\figwidths}{0.45\columnwidth}

\begin{figure}[tb]
 \centering
 \resizebox{\columnwidth}{!} {\includegraphics[angle=270]{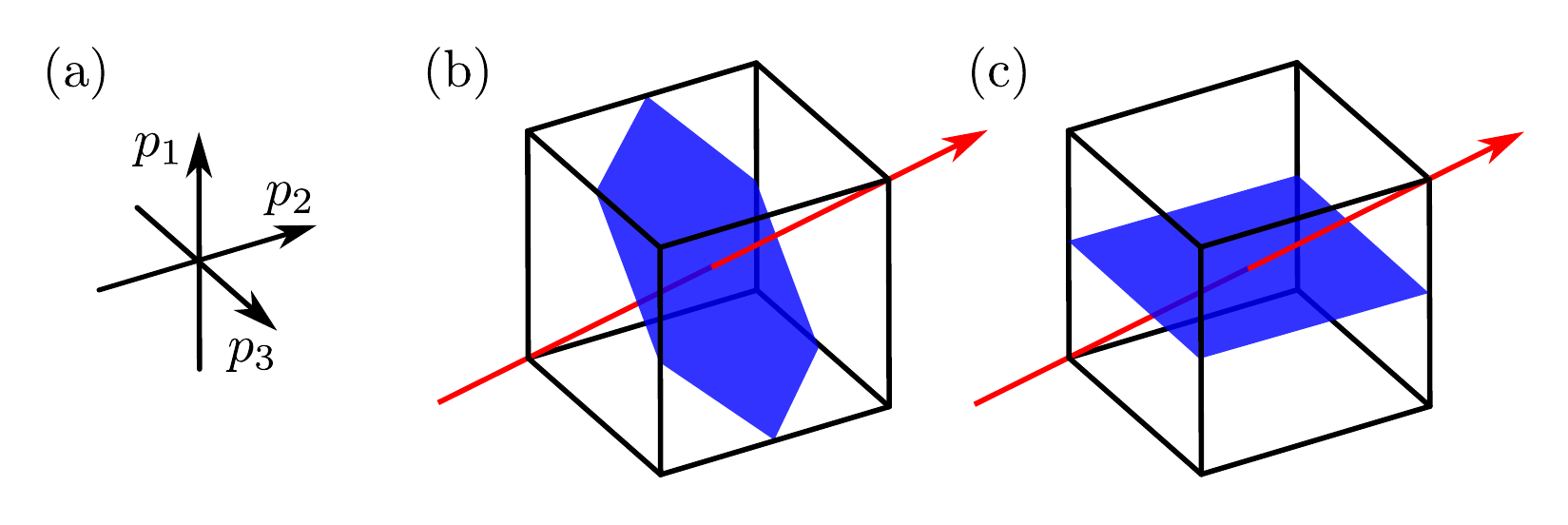}}
 \caption{Representation of the momentum integrals in \eqnn{eq:longtana}{boodles} for $N=3$. The axes (a) are positioned such that the origin is at the centre of each of the cubes, which are cut by (b) the centroid dividing surface $h({\ov p}_0)$ (blue), and (c) the dividing surface $h(p_1)$ (blue). The red arrow represents the centroid axis. This picture can be generalized to $N>3$, by replacing the cubes with $N$-dimensional hypercubes.}
\end{figure}
  
 The only formal properties  of hypercubes that we need are, first that a hypercube has $2^N$ vertices, second that one can represent the hypercube by constructing a graph showing the connections between its vertices, and third that the graph for a hypercube of dimension $N$ can be made by connecting equivalent vertices on the graphs of two hypercubes of dimension $N-1$. Figure 2 illustrates this last point, showing how the graph for a cube ($N=3$) can be made by connecting equivalent vertices on the graphs for two squares ($N=2$). Figure 2 also introduces the (self-evident) notation that we will use to label vertices; e.g. $(-1,1,1)$ refers to the vertex on an $N=3$ hypercube (i.e.\ a cube) located
 at $p_1=-p_{\rm max}$, $p_2=p_3=p_{\rm max}$.
 
  \begin{figure}[b]
 \resizebox{\columnwidth}{!} {\includegraphics[angle=270]{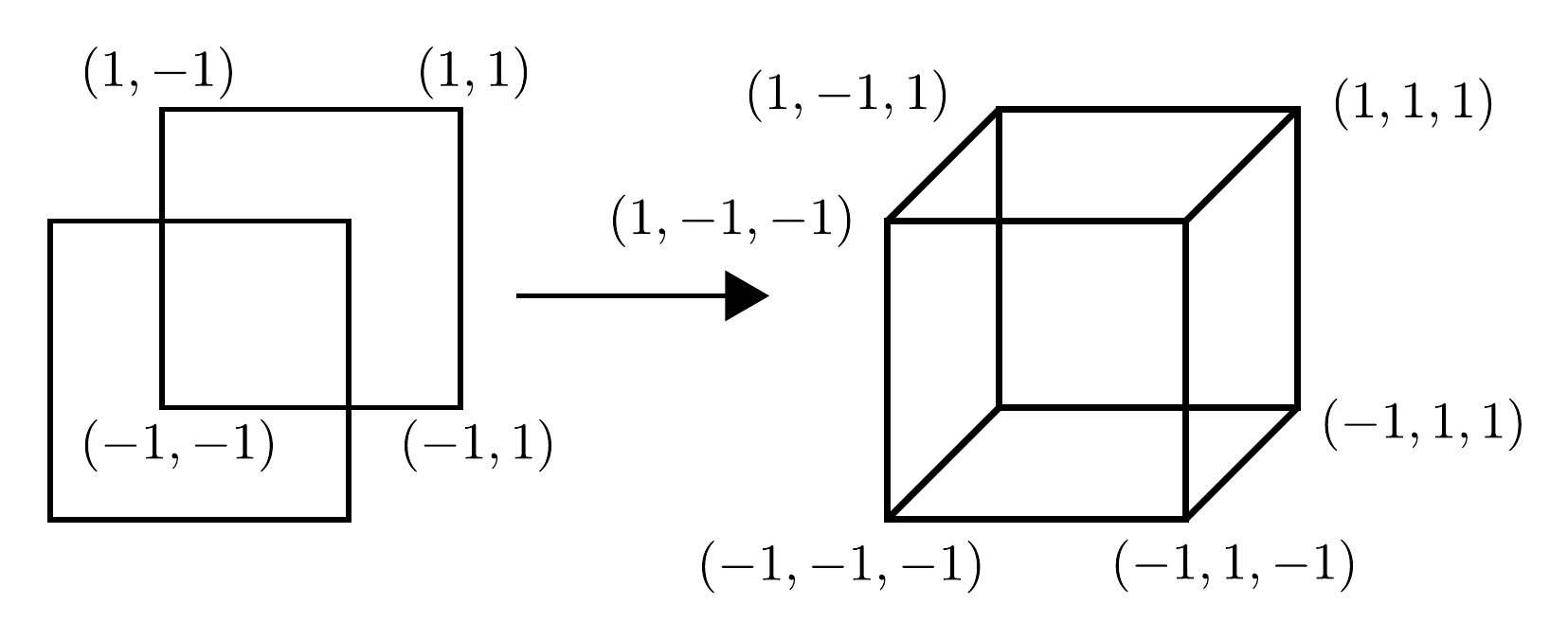}}
 \caption{Diagram showing how a cube can be built up by connecting  the equivalent vertices on two  squares. One can similarly build up an $N$-dimensional hypercube by connecting the equivalent vertices on two $(N\!-\!1)$-dimensional hypercubes. This figure also illustrates the notation used in the text to label the vertices of a hypercube.}
\end{figure}
 
These properties  allow one to build up a hypercube by adding together its {\em subcubes} in a recursive sequence.  By subcube we mean that each $p_i$ is confined to either the positive or negative axis; there are therefore
 $2^N$ subcubes, each corresponding to a different vertex of the hypercube (so we can label the subcubes using the vertex notation introduced above).  Figure 3 shows how one can build up an $N=3$ hypercube (i.e.\ a cube) by adding its subcubes together recursively, joining first two individual subcubes along a line, then joining two lines
 of subcubes in the form of a square, and finally joining two squares of subcubes to give the entire cube.  The analogous sequence can be used to build up a hypercube of any dimension $N$ from its subcubes, and will be useful in Sec.~IV.B.
 
 \begin{figure}[tb]
 \resizebox{.8\columnwidth}{!} {\includegraphics{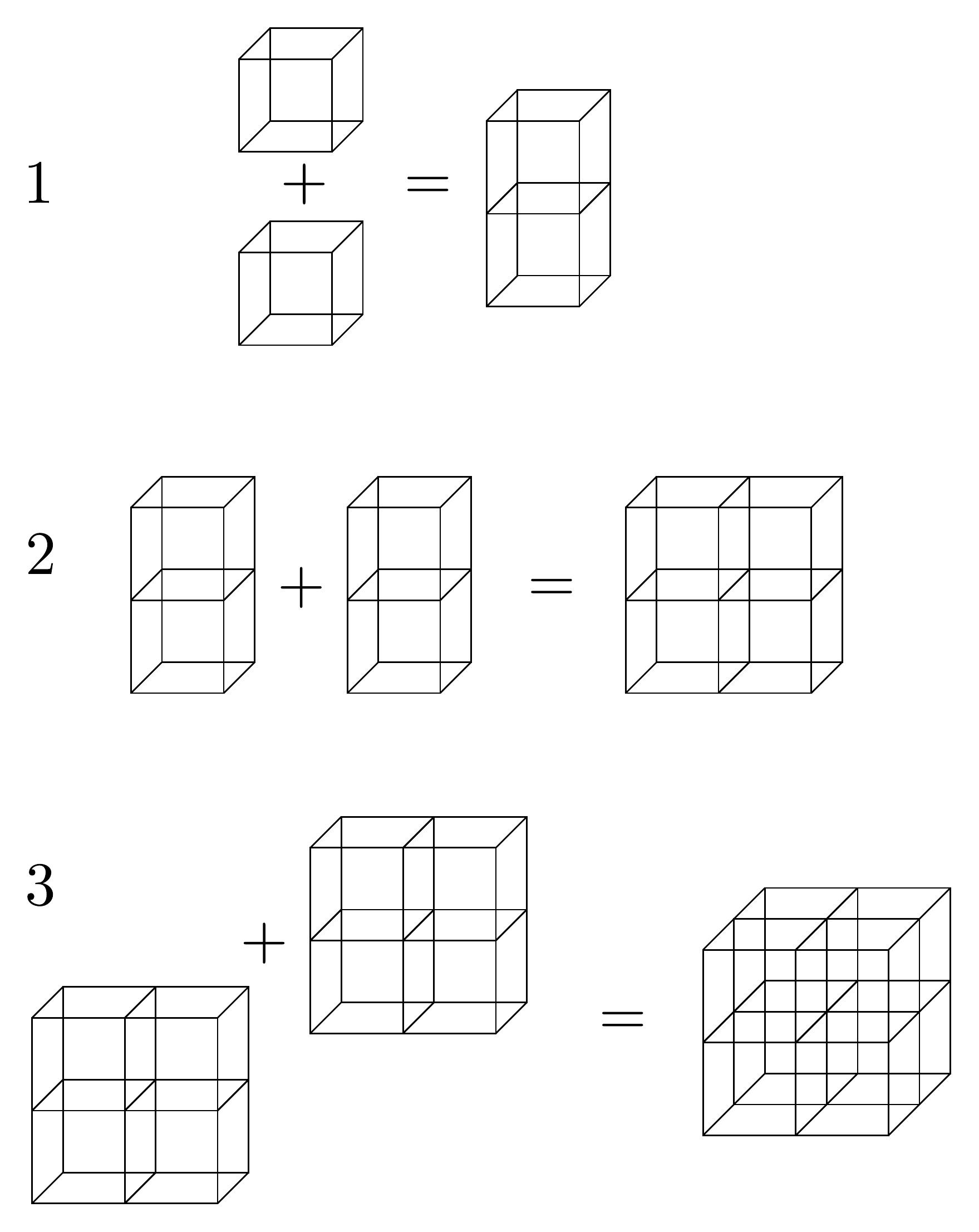}}
 \caption{Diagram showing how a cube can be built up recursively in three steps from its eight subcubes. One can similarly build up an $N$-dimensional hypercube in $N$ steps from its $2^N$ subcubes.}
\end{figure}
 
  We now define the energies
  \begin{align}
E_i\equiv E^-(p_i)&={p_i^2\over 2m}+V_{\rm prod}&p_i>0\nonumber\\
&={{p}_i^2\over 2m}+V_{\rm reac}&p_i<0\label{nofood}
 \end{align} 
 and introduce the notation ${\w p}_i$, such that
 \begin{align}
 {\w p}_i&=-\sqrt{p_i^2+2m(V_{\rm prod}-V_{\rm reac})} &p_i>0\nonumber\\
 {\w p}_i&=+\sqrt{p_i^2+2m(V_{\rm reac}-V_{\rm prod})} &p_i<0\label{mister}
  \end{align} 
where $V_{\rm reac}$ and $V_{\rm prod}$ are the asymptotes of the potential $V(x)$ in the reactant ($x\to-\infty$) and product ($x\to\infty$) regions;  i.e.\ the tilde has the effect of converting a product momentum to the reactant momentum corresponding to the same energy $E_i$, and vice versa. Note that we will not need to interconvert between the reactant and product momenta if one or other of them is imaginary, and hence the square roots in \eqn{mister} are always real. 

For a symmetric barrier, it is clear that ${\w p}_i=-p_i$, and from this it is easy to show  that 
  \begin{align}
 A_N({\w{\bf p}}) = - A_N({\bf p}) \ \ \ \ \ \ \text{for symmetric barriers}
 \label{goody}
  \end{align}
 where ${\w{\bf p}}\equiv({\w p}_1,{\w p}_2,\dots,{\w p}_N)$; i.e.\ 
 $ A_N({{\bf p}})$ is antisymmetric with respect to inversion through the origin. 
 Clearly this antisymmetry ensures that the integration of $ A_N({\w{\bf p}})$ over the entire hypercube
  (i.e.\ with the side function omitted) gives zero. This integral is also zero for an asymmetric barrier, but there is then no simple cancellation of $ A_N({{\bf p}})$ with $ A_N({\w{\bf p}})$. 
  
Finally, we note that $A_N({{\bf p}})$ is symmetric with respect to cyclic permutations of the $p_i$, and thus has an $N$-fold axis of rotational symmetry around the diagonal of the hypercube on which all $p_i$ are equal. We will refer to this diagonal as the `centroid axis', since  displacement along this axis measures the displacement of the momentum centroid ${\overline p}_0=\sum_{i=1}^Np_i/N$.

 \subsection{Ring-polymer flux-side time-correlation function}
 
 It is straightforward to modify the above derivation to obtain the $t\to \infty$ limit of the ring-polymer flux-side time-correlation function 
 ${ C}_{\rm fs}^{[N]}(t)$. The only change necessary is to replace the side function $h(z_1)$ by $h[f({\bf z})]$, which gives
 \begin{align}
\lim_{t\rightarrow \infty} { C}_{\rm fs}^{[N]}(t)=\int\! d{\bf p}\,A_N({\bf p})h[{\overline{f}}(\bf p)]
\label{boodles} 
 \end{align}
where $ A_N({{\bf p}})$ is defined in \eqn{pete}, and ${\overline{f}}(\bf p)$ is defined by 
 \begin{align}
 \lim_{t\rightarrow \infty} h[f({\bf p}t/m)]=h[{\overline f}({\bf p})]\label{lunchtime}
 \end{align} 
i.e.\ ${\overline{f}}(\bf p)$ is the limit of ${{f}}(\bf p)$ at very large distances. In the special case that
  ${{f}}({\bf q})={\overline q}_0$,  we obtain ${\overline{f}}({\bf p})={\overline p}_0={{f}}({\bf p})$;
   but in general ${\overline{f}}({\bf p})\ne {{f}}(\bf p)$. A time-independent limit of \eqn{lunchtime}
is guaranteed to exist, since otherwise ${{f}}({\bf q})$ would not satisfy the requirements of a dividing surface.
 
 Whatever the choice of ${{f}}({\bf q})$, it is clear that the (permutationally invariant) $h[{\overline{f}}(\bf p)]$ encloses a different part of the hypercube than does $h(p_1)$. For example, if ${{f}}({\bf q})={\ov q}_0$ and $N=3$, then $h[{\overline{f}}({\bf p})]=h({\ov p}_0)$ cuts out the half of the cube on the positive side of the hexagonal cross-section shown in Fig.~1b, whereas $h(p_1)$ cuts off the top half of the cube on the $p_1$ axis (Fig.~1c). Thus we cannot in general expect the $t\to \infty$ limits of ${ C}_{\rm fs}^{[N]}(t)$ and ${\ov C}_{\rm fs}^{[N]}(t)$ to be the same, unless $A_N({\bf p})$ satisfies some special properties in addition to those just mentioned. We will show in the next two Sections that $A_N({\bf p})$ does satisfy such properties if there is no recrossing of any surface orthogonal to ${{f}}({\bf q})$ in ring-polymer space.
  
\section{Ring-polymer momentum integrals}

\subsection{Structure of $A_N({\bf p})$}
One can show using scattering theory (see Appendix B) that $A_N({\bf p})$ consists of the terms
\begin{align}
A_N({\bf p}) = a_N({\bf p})\left[\prod_{i=1}^{N-1}\delta(E_{i+1}-E_i)\right]+r_N({\bf p})
\label{piggy}
 \end{align}
where $a_N({\bf p})$ is some function of ${\bf p}$, and $r_N({\bf p})$ satisfies 
\begin{align}
r_N(p_1,\dots,{\w p}_j,\dots,p_N)=-\left|{\w p}_j\over p_j\right| r_N(p_1,\dots,p_j,\dots,p_N)
\label{flap}
\end{align}
(where the dots indicate that all the $p_i$ except $p_j$ take the same values on both sides of the equation). \Eqn{flap} is equivalent to stating that  $r_N({\bf p})$ alternates in sign between {\em adjacent subcubes} (i.e.~subcubes that differ in respect of just one axis), or that $r_N({\bf p})$ takes opposite signs in {\em even} and {\em odd} subcubes (where a subcube is defined to be even/odd  if it has an even/odd number of axes for which $p_i<0$). Note that $r_N({\bf p})=0$ if any ${\w p}_i$, $i=1\dots N$, is imaginary (see Appendix B).

The first term in \eqn{piggy} describes a set of $2^N$ $\delta$-function spikes running along all the lines in the hypercube for which the energies $E_i$, $i=1\dots N$, are equal. 
There is one such line in every subcube. Two of these lines point in positive and negative directions along the centroid axis (i.e.\ the diagonal of the hypercube). The other $2^N-2$ {\em off-diagonal} spikes radiate out from this axis. If the barrier is symmetric, then each off-diagonal
spike is a straight line joining the centre of the hypercube to one of its vertices. If the barrier is asymmetric, the off-diagonal spikes are hyperbolae [on account of \eqn{mister}]. The off-diagonal spikes are distributed with $N$-fold rotational symmetry about the centroid axis because of the invariance of $A_N({\bf p})$ under cyclic permutations; e.g.\ for $N=3$, the spikes $(-1,1,1)$, $(1,-1,1)$, $(1,1,-1)$ (where this notation identifies each spike by the subcube that it runs through)  rotate into one another under cyclic permutation of the beads; see Fig.~4.

\begin{figure}[tb]
 \resizebox{.6\columnwidth}{!} {\includegraphics{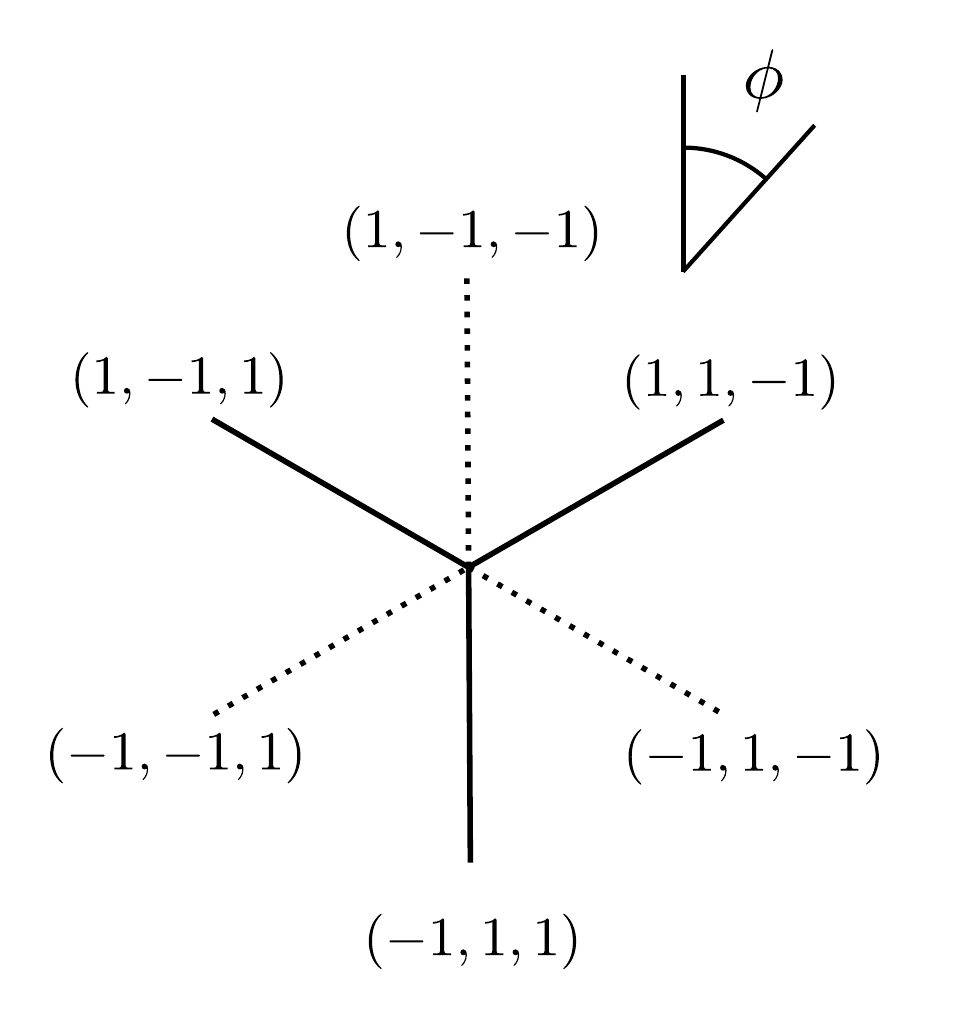}}
 \caption{Plot of the off-diagonal spikes in $A_N({\bf p})$ for $N=3$, obtained by looking down the centroid axis (the red arrow in Fig.~1b).}
\end{figure}

\subsection{Cancellation of the term $r_N({\bf p})$}

We now show that $r_N({\bf p})$ in \eqn{piggy} contributes zero to ${\ov C}_{\rm fs}^{[N]}(t)$ and ${ C}_{\rm fs}^{[N]}(t)$ in the limits $t,N\to\infty$, and may therefore be ignored when discussing whether ${ C}_{\rm fs}^{[N]}(t)$ gives the exact quantum rate in these limits. This property is easy to show for a symmetric barrier, for which \eqnn{goody}{flap} imply that $r_N({\bf p})$ is zero for all even $N$, and thus that the contribution to the integral from $r_N({\bf p})$ tends to zero in the limit \ntya. For an asymmetric barrier, $r_N({\bf p})$ is in general non-zero. However, we now show that the alternation in sign between adjacent subcubes [\eqn{flap}] causes $r_N({\bf p})$ to cancel out in both ${\ov C}_{\rm fs}^{[N]}(t)$  and ${ C}_{\rm fs}^{[N]}(t)$ in the limits $t,N\to\infty$.
 
 This cancellation is easy to demonstrate for  ${\ov C}_{\rm fs}^{[N]}(t)$: One simply notes that the side-function $h(p_1)$ encloses an even number of subcubes, which can be added together in adjacent pairs. For example, if we add together the adjacent subcubes $(1,\dots,1,1)$ and $(1,\dots,1,-1)$  (where the dots indicate that the intevening values of 1 and $-1$ are the same for the two subcubes), we obtain
  \begin{align}
 &\int_{0}^\infty\! dp_1\dots\int_{0}^\infty\! dp_{N-1}\int_{0}^\infty\! dp_{N}\,r_N({\bf p})h(p_1)\no\\+&\int_{0}^\infty\! dp_1\dots\int_{0}^\infty\! dp_{N-1}\int_{-\infty}^0\! dp_{N}\,r_N({\bf p})h(p_1)\label{yokel}
 \end{align}
 (where the dots indicate that the integration ranges for $p_i$, $i=2\dots N\!-\!2$ are the same in both terms). 
We can  change the limits on the last integrand to $0\rightarrow\infty$ by transforming the integration variable from $p_N$ to ${\w p}_N$, and using the relation $p_idp_i={\w p}_id{\w p}_i$ [see \eqn{mister}]. \Eqn{flap} then ensures that the two terms in \eqn{yokel} cancel out. Hence the contribution from $r_N({\bf p})$ cancels out in the \longt limit of ${\ov C}_{\rm fs}^{[N]}(t)$ (for any $N>0$). 
  
Using similar reasoning, we can show that the contribution from $r_N({\bf p})$ to
 ${C}_{\rm fs}^{[N]}(t)$  cancels out in the limits $t,N\to\infty$. For finite $N$, this cancellation  is in general\cite{centy} only partial, because the function $h[{\overline f}({\bf p})]$ encloses different volumes in any two adjacent subcubes. However, one can show that the total mismatch in the volumes  enclosed in the even subcubes and the odd subcubes tends rapidly to zero as \ntya. The trick is to  build up the hypercube recursively, by extending to higher $N$ the sequence shown in Fig.~3 for $N=3$. The $j$th step in this sequence can be written
  \begin{align}
S(N)=&\int_{-\infty}^\infty\! dp_1\dots\int_{-\infty}^\infty\! dp_{j}\int_{0}^\infty\! dp_{j+1}\dots\int_{0}^\infty\! dp_{N} \nonumber \\
& \times \,r_N({\bf p})h[{\overline{f}}(\bf p)]\no\\
=&\int_{-\infty}^\infty\! dp_1\dots\int_{-\infty}^\infty\! dp_{j-1}\int_{0}^\infty\! dp_{j}\dots\int_{0}^\infty\! dp_{N}\,\nonumber \\
& \times r_N({\bf p})h[{\overline{f}}(\bf p)]\no\\
+&\int_{-\infty}^\infty\! dp_1\dots\int_{-\infty}^\infty\! dp_{j-1}\int_{-\infty}^0\! dp_{j} \nonumber \\
& \times \int_{0}^\infty\! dp_{j+1}\dots\int_{0}^\infty\! dp_{N}\,r_N({\bf p})h[{\overline{f}}(\bf p)]
\label{ladder}
\end{align}
(where the first set of dots in each term indicates that the intervening integration ranges are $-\infty<p_i<\infty$, and the second set that they are  $0<p_i<\infty$). Because each subcube in the second term is adjacent to its counterpart in the third term, there is an almost complete cancellation in the $r_N({\bf p})$ terms. All that is left is the residue,
\begin{align}
S(N)
=&\int_{-\infty}^\infty\! dp_1\dots\int_{-\infty}^\infty\! dp_{j-1}\int_{0}^\infty\! dp_{j}\dots\int_{0}^\infty\! dp_{N}\,r_N({\bf p})\no\\
&\times\left\{ h[{\overline{f}}(p_1,\dots,p_j,\dots,p_N)] \no \right.\\
&\qquad\left. - h[{\overline{f}}(p_1,\dots,{\w p}_j,\dots,p_N)]\right\}
\label{residue}
\end{align}
which occupies the volume sandwiched between the two heaviside functions. Appendix C shows that this volume is a thin strip on the order of $N$ times smaller than the volume occupied by $r_N({\bf p})$
 in each of the two terms that were added together in \eqn{ladder}. Now, each of these terms was itself the result of a similar addition in the $j-1$ th step, which also reduced the volume occupied by $r_N({\bf p})$ by a factor on the order of $N$, and so on. As a result, the volume occupied by $r_N({\bf p})$ after the $N$th (i.e.\ final) step is on the order of $N^N$ times smaller than the volume of a single subcube. The mismatch in volume between the even and odd subcubes thus tends rapidly to zero in the limit \ntya, with the result that $r_N({\bf p})$ cancels out completely \cite{unless} in ${ C}_{\rm fs}^{[N]}(t)$ in the limits $t,N\to\infty$.

\subsection{Comparison of $\delta$-function spikes}

We have just shown that only the first term in \eqn{piggy} contributes to 
 ${\ov C}_{\rm fs}^{[N]}(t)$  and ${ C}_{\rm fs}^{[N]}(t)$ in the limits $t,N\to\infty$.   Any difference between these quantities can thus be accounted for by comparing which spikes are enclosed by the side functions $h(p_1)$ and $h[{\overline f}({\bf p})]$. It is clear that both $h(p_1)$ and $h({\ov p}_0)$ enclose the spike that runs along the centroid axis in a positive direction, and exclude the spike that runs in a negative direction. A little thought shows that this property must hold for any choice of $h[{\overline f}({\bf p})]$ (since the positive spike corresponds to all momenta $p_i$ travelling in the product direction as \longta, and vice versa for the negative spike).

  Any difference between the $t,N\to\infty$ limits of ${\ov C}_{\rm fs}^{[N]}(t)$ and ${ C}_{\rm fs}^{[N]}(t)$ can therefore be explained in terms of which off-diagonal spikes are enclosed by $h(p_1)$ and $h[{\overline f}({\bf p})]$. These functions will enclose different sets of spikes. For example, for a symmetric barrier, with $N=3$, the function $h({\ov p}_0)$ encloses the off-diagonal spikes $(-1,1,1)$, $(1,-1,1)$ and $(1,1,-1)$, whereas $h(p_1)$ encloses $(1,-1,1)$, $(1,1,-1)$ and $(1,-1,-1)$.
 
We have therefore obtained the result that the $t,N\to\infty$ limit of ${ C}_{\rm fs}^{[N]}(t)$ is identical to that of ${\ov C}_{\rm fs}^{[N]}(t)$ (and thus gives the exact quantum rate) {\em if} the contribution from each off-diagonal spike to $A_N({\bf p})$ is individually zero. 
 We make use of this important result in the next Section.

\section{Effects of recrossing}

The results just obtained show that quantum TST will give the exact quantum rate if two conditions are satisfied. First, there must be no recrossing of the cyclically invariant dividing surface $f({\bf q})$ (by which we mean simply that ${ C}_{\rm fs}^{[N]}(t)$ is time-independent). Second, each of the off-diagonal spikes [in the first term of \eqn{piggy}] must contribute zero to ${ C}_{\rm fs}^{[N]}(t)$ in the long-time limit. We now show that this last condition is satisfied if there is no recrossing of any dividing surface orthogonal to $f({\bf q})$ in ring-polymer space. 

\subsection{Orthogonal dividing surfaces}

A dividing surface $g({\bf q})$ orthogonal to $f({\bf q})$ satisfies
\begin{align}
\sum_{i=1}^N  {\partial g({\bf q})\over \partial q_i}  {\partial f({\bf q})\over \partial q_i} &=0\label{orthy}
\end{align}
When $f({\bf q})={\overline q}_0$, the surface $g({\bf q})$ can be any function of any linear combination of polymer beads orthogonal to ${\overline q}_0$.
For a more general (cyclically permutable) $f({\bf q})$, $g({\bf q})$ will also take this form close to the centroid axis (where, by definition, all degrees of freedom orthogonal to the centroid vanish), and will assume a more general curvilinear form away from this axis.

 By no recrossing of $g({\bf q})$, we mean that the time-correlation function
 \begin{align}
M_{\rm fs}^{[N]}(t)=&\int\! d{\bf q}\, \int\! d{\bf z}\,\int\! d{\bf \Delta}\,{\cal \hat F}[f({\bf q})]h[g({\bf z})]
\nonumber\\ 
&\times\prod_{i=1}^{N}\expect{q_{i-1}-\Delta_{i-1}/2|e^{-\beta_N{\hat H}}|q_i+\Delta_i/2}
\nonumber\\ 
&\qquad\times \expect{q_i+\Delta_i/2|e^{i{\hat H}t/\hbar}|z_i} \no\\
& \qquad\times \expect{z_i|e^{-i{\hat H}t/\hbar}|q_i-\Delta_i/2}
\label{puss} 
\end{align}
is time-independent. We know from Part I that the \shortt limit of $M_{\rm fs}^{[N]}(t)$ is zero, since the flux and side dividing surfaces are different. Hence no recrossing of $g({\bf q})$ implies that $M_{\rm fs}^{[N]}(t)$ is zero for all time $t$, indicating that there is no net passage of flux from the initial distribution on $f({\bf q})$ through the surface $g({\bf q})$. Taking the \longt limit
 (using the same approach as in Sec.~III), we obtain 
 \begin{align}
\lim_{t\rightarrow \infty} { M}_{\rm fs}^{[N]}(t)&=\int\! d{\bf p}\,A_N({\bf p})h[{\overline{g}}(\bf p)]\no\\
&=0\ \ \ \ \ \ {\text{if no recrossing of }}g({\bf q})
\label{lunghi} 
 \end{align}
 where $A_N({\bf p})$ is defined in \eqn{pete}, and 
${\overline g}({\bf p})$ is defined analogously to ${\overline f}({\bf p})$, i.e.
 \begin{align}
 \lim_{t\rightarrow \infty} h[g({\bf p}t/m)]=h[{\overline g}({\bf p})]
 \label{baffi}
 \end{align} 
In the \nty limit, the contribution of $r_N({\bf p})$ to ${ M}_{\rm fs}^{[N]}(t)$ cancels out  (for the same reason that it cancels out in ${ C}_{\rm fs}^{[N]}(t)$---see Sec.~IV.B).  \Eqn{lunghi} is thus equivalent to stating that the total contribution to $A_N({\bf p})$ from the spikes  enclosed by $h[{\overline{g}}(\bf p)]$ is zero if there is no recrossing of $g({\bf q})$.

\subsection{Effect of no recrossing orthogonal to $f({\bf q})$}

If there is no recrossing of any ${g}({\bf q})$ orthogonal to $f({\bf q})$, we can use \eqn{lunghi} to generate a set of equations giving constraints on the spikes.
Let us see what effect these constraints have in the simple case that $N=3$ and $f({\bf q})={\ov q}_0$. \cite{cancel} We can construct dividing surfaces $g({\bf q})$ orthogonal to $f({\bf q})$ by taking any function of the normal mode coordinates
\begin{align}
Q_{x}&={1\over\sqrt{6}}\left(2q_1-q_2-q_3\right)\no\\
Q_{y}&={1\over\sqrt{2}}\left(q_2-q_3\right)
\end{align}
Let us take 
\begin{align}
g_r({\bf q})&=\sqrt{Q_{x}^2+Q_y^2}-r^\ddag\no\\
g_F({\bf q}) &= F\!\left[\phi(Q_x,Q_y) \right]
\end{align}
where $r^\ddag>0$ specifies the position of surface $g_r({\bf q})$, and $F$ can be chosen to be any smooth function \cite{smooth} of the angle
\begin{align}
\phi(Q_x,Q_y)=\arctan(Q_y/Q_x)
\end{align}
Clearly  $g_r({\bf q})$ and $\phi$ are  polar coordinates in the plane orthogonal to the centroid axis. If there is no recrossing of 
$g_r({\bf q})$ or $g_F({\bf q})$, then \eqn{lunghi} will hold with
 \begin{align}
{\overline g}_r({\bf p})&=\lim_{\epsilon\rightarrow 0}\sqrt{P_{x}^2+P_y^2}-\epsilon\no\\
{\overline g}_F({\bf p})&=F\!\left[\phi(P_x,P_y) \right]
\end{align}
in place of ${\overline g}({\bf p})$ [where  $(P_x,P_y)$ are the combinations of $p_i$ analogous to $(Q_x,Q_y)$]. Now, ${\overline g}_r({\bf p})$  is a thin cylinder enclosing the centroid axis, and hence this function gives the constraint that the contributions to $A_N({\bf p})$ from the two spikes lying along this axis (in positive and negative directions) cancel out.\cite{symmetric} We are then free to choose $F$ so that $h[{\overline g}_F({\bf p})]$ encloses each off-diagonal spike in turn, since no two off-diagonal spikes pass through the same angle $\phi$ (see Fig.~4). We do not need to worry about the spikes along the centroid axis (which appear as a point at the origin---see Fig.~4), since we have just shown that they cancel out. \Eqn{lunghi} then gives a set of constraints, each of which specifies that the contribution to $A_N({\bf p})$ from one of the spikes is individually zero [if there is no recrossing orthogonal to  $f({\bf q})$].

 In Appendix D, we show that this result  generalizes to any $N$ and to any  choice of the cyclically invariant dividing surface
 $f({\bf q})$. The $t,N\to\infty$ limit of ${ C}_{\rm fs}^{[N]}(t)$ is therefore equal to the \longt limit of ${\ov C}_{\rm fs}^{[N]}(t)$ if there is no recrossing orthogonal to $f({\bf q})$. Since the \shortt limit of ${ C}_{\rm fs}^{[N]}(t)$ is by definition equal to its \longt limit if there is also no recrossing of $f({\bf q})$, we have therefore derived
 the main result of this article: {\em quantum TST (i.e.\ RPMD-TST) gives the exact quantum rate for a one-dimensional system
  if there is no recrossing of $f({\bf q})$, nor of any surface orthogonal to it in ring-polymer space}. We will show in Sec.~VI that this result generalises straightforwardly to multi-dimensions.
 
\subsection{Interpretation}

Quantum TST therefore differs from classical TST in requiring an extra condition to be satisfied if it is to give the exact rate:  in addition to no recrossing of the dividing-surface
$f({\bf q})$, there should also be no
recrossing (by the exact quantum dynamics) of surfaces in the ($N\!-\!1$)-dimensional 
space orthogonal to $f({\bf q})$. In the limit $t\to 0_+$, this space describes  {\em fluctuations} in the positions of the ring-polymer beads.
The extra condition is therefore satisfied automatically in the classical (i.e.\ high temperature) limit, where it is impossible to recross any surface orthogonal to $f({\bf q})$, since the initial distribution of polymer beads is localised at a point and only the projection of the momentum along the centroid axis is non-zero.
For similar reasons, it is also impossible to recross any surface orthogonal to $f({\bf q})={\overline q}_0-q^\ddag$ for a parabolic barrier 
at any  temperature (at which the parabolic-barrier rate is defined).
As a result, quantum TST gives the exact rate in the classical limit and for a parabolic barrier, provided there is no recrossing of $f({\bf q})$ (which condition is satisfied
for a parabolic barrier when $q^\ddag$ is located at the top of the barrier).

In a real system, there will always be some recrossing of surfaces orthogonal to $f({\bf q})$ on account of the anharmonicity. 
However, the amount of such recrossing is zero in the high temperature limit (see above), and will only become significant at temperatures  
sufficiently low that the \shortt distribution of polymer beads is delocalised beyond the parabolic tip of the potential barrier. In practice, this means that
 quantum TST (i.e.\ RPMD-TST) will give a good approximation to the exact quantum rate at temperatures above the cross-over to deep-tunnelling (provided the reaction is not dominated by dynamical recrossing or real-time coherence effects). On reducing the temperature below cross-over, the amount of recrossing orthogonal to $f({\bf q})$ will increase, with the result that quantum TST will become progressively less accurate. Previous work on RPMD \cite{jor,rates,refined,bimolec,ch4,anrev}
  and related instanton methods \cite{jor,bill,cole,benders,jonss,spanish,kastner1,kastner2,equiv} has shown that this deterioration in accuracy is gradual, with the RPMD-TST rate typically giving a good approximation to the exact quantum rate at temperatures down to half the cross-over temperature and below. 
   
\subsection{Correction terms}

An alternative way of formulating the above is to regard the ${ M}_{\rm fs}^{[N]}(t)$ as a set of correction terms, which can be added to 
 ${ C}_{\rm fs}^{[N]}(t)$ in order to recover the exact quantum rate in the limits $t\to\infty$. The orthogonal
  surfaces $g({\bf q})$ should be chosen such that  the resulting sum of terms contains the same set of spikes in the \longt limit as does ${\ov C}_{\rm fs}^{[N]}(t)$. For example, if $N=3$ and $f({\bf q})={\ov q}_0$, we can define two time-correlation functions ${M}_{1}(t)$ and ${M}_{2}(t)$ which use dividing surfaces of the form of $g_F({\bf q})$, with $F$ chosen to enclose, respectively, the spikes $(1,1,-1)$ and $(1,-1,-1)$. The corrected flux-side time-correlation function
\begin{align}
{ C}_{\rm corr}^{[N=3]}(t) = { C}_{\rm fs}^{[N=3]}(t) - {M}_{1}(t) + {M}_{2}(t)
\end{align}
then contains the same spikes in the \longt limit as ${\ov C}_{\rm fs}^{[N]}(t)$. Since ${M}_{1}(t)$ and ${M}_{2}(t)$ are zero in the limit \shortta, it follows that ${ C}_{\rm corr}^{[N=3]}(t)$  interpolates between the RPMD-TST rate in the limit \shortta, and the exact quantum rate in the limit \longta. \cite{cancel} Clearly ${M}_{1}(t)$ and ${M}_{2}(t)$ will be zero for all values of $t$ if there is no recrossing of surfaces orthogonal to $f({\bf q})$ in ring-polymer space. This treatment generalizes in an obvious way to $N>3$.
 An alternative way of stating the result of Sec.~V.B is thus that  ${ C}_{\rm fs}^{[N]}(t)$ gives the exact rate in the \longt limit when added to correction terms which are zero in the absence of recrossing. 
 
\section{Application to multi-dimensional systems}

Here we outline the modifications needed to extend Secs.~III-V to multi-dimensional systems. As in Secs.~III-V,
we make use of quantum scattering theory, but we emphasise that the results obtained here apply also
in the condensed phase, provided there is the usual separation in timescales between barrier-crossing and equilibration.\cite{isomer}

Following Part I, we represent the space of an $F$-dimensional reactive scattering system using cartesian coordinates $q_j$, $j=1\dots F$, and define ring-polymer coordinates ${\bf q}\equiv \{{\bf q}_{1},\dots, {\bf q}_{N}\}$,
where  ${\bf q}_i\equiv \{q_{i,1},\dots,q_{i,F}\}$ is the geometry of the $i$th replica of the system.
Analogous generalizations can be made of $\bf z$, $\bf p$, $\bf \Delta$, and so on. We then construct a multi-dimensional
 version of $C_{\rm fs}^{[N]}(t)$  by 
making the replacements
  \begin{align}
\ket{q_i+\Delta_i/2}\rightarrow \ket{q_{i,1}+\Delta_{i,1}/2,\dots ,q_{i,F}+\Delta_{i,F}/2 }\label{tires}
\end{align}
in \eqn{utter}, and integrating over the multi-dimensional coordinates $({\bf q},{\bf z},{\bf \Delta})$.  The dividing surface $f({\bf q})$ is now invariant under {\em collective} cyclic permutations of the coordinates ${\bf q}$, and is thus a permutationally invariant function of the replicas ${\ssi}\equiv \{\sigma_1({\bf q}_{1}),\dots,\sigma_N({\bf q}_{N})\}$ of a (classical) reaction coordinate $\sigma(q_1,\dots,q_F)$. 

It is straightforward to analyze the \longt behaviour of $C_{\rm fs}^{[N]}(t)$ by combining the analysis of Secs.~III-V with centre-of-mass-frame scattering theory. All we need to note is that the relative motion of the reactant or product molecules can be described by a one-dimensional scattering coordinate, with all other degrees of freedom being described by channel functions \cite{taylor}  (which include the rovibrational states of the scattered molecules, and also specify whether the system is in the reactant or product arrangement). We will denote the momentum of the $i$th replica along the scattering coordinate as $\pi_i$, using the convention that $\pi_i$ is negative in the reactant
 arrangement and positive in the product arrangement. Since all other internal degrees of freedom are bound, it follows that
  \begin{align}
\lim_{t\rightarrow \infty} h[\sigma_i({\bf p}_it/m)]=h(\pi_i)
 \end{align}
 
This last result allows us to construct a multi-dimensional generalisation of the hybrid function ${\ov C}_{\rm fs}^{[N]}(t)$
 by replacing $h[f({\bf q})]$ in $C_{\rm fs}^{[N]}(t)$ by $h[\sigma_i({\bf q}_i)]$. One can show (by generalizing Appendix~A) that the multi-dimensional
 ${\ov C}_{\rm fs}^{[N]}(t)$ gives the exact quantum rate in the limit \longta. We then take the \longt
  limits of ${\ov C}_{\rm fs}^{[N]}(t)$ and $C_{\rm fs}^{[N]}(t)$ by using the  scattering relation
\begin{align}
\hat\Omega_-\ket{\pi_i}\ket{v_i}&= \ket{\phi^-_{{\pi}_i,v_i}}
\end{align}
where ${\hat \Omega}_-$ is the (multi-dimensional) M{\o }ller operator,\cite{taylor} $\ket{\pi_i}$ is a momentum eigenstate, $\ket{v_i}$ is a reactant or product channel function, and $\ket{\phi^-_{{\pi}_i,v_i}}$ is a scattering eigenstate satisfying outgoing boundary conditions. As in one-dimension, we obtain integrals over an $N$-dimensional hypercube:
 \begin{align}
\lim_{t\rightarrow \infty} {\ov C}_{\rm fs}^{[N]}(t)&=\int\! d{\ppi}\,A_N({ \ppi})h(\pi_i)\\
\lim_{t\rightarrow \infty} { C}_{\rm fs}^{[N]}(t)&=\int\! d{\ppi}\,A_N({ \ppi}) h[{\overline{f}}(\ppi)]\
\label{eq:longtnf} 
 \end{align}
  where $\ppi\equiv\{\pi_1,\dots,\pi_N\}$, and $A_N({\ppi})$  is a generalisation of $A_N({\bf p})$, obtained by making the replacements of \eqn{tires} in 
\eqn{pete}, replacing $\ket{\phi^-_{p_i}}$ by $\ket{\phi^-_{{\pi}_i,v_i}} $,  and summing over $v_i$. The function ${\overline f}({\ppi})$ is a multi-dimensional generalisation of ${\overline f}({\bf p})$, and satisfies
  \begin{align}
 \lim_{t\rightarrow \infty} h[f({\bf p}t/m)]=h[{\overline f}({\ppi})]\label{basingstoke}
 \end{align} 
(which is equivalent to stating that $f({\bf q})$ separates cleanly the reactants from the products in the limit \longta). 


The derivation of Appendix B generalizes straightforwardly to multi-dimensions,
 with the result that $A_N({\ppi})$ has the analogous structure to $A_N({\bf p})$ in \eqn{piggy}. Following Sec.~IV and Appendix C, one can show  that only the $\delta$-function spikes [corresponding to the first term in \eqn{piggy}] contribute to ${\ov C}_{\rm fs}^{[N]}(t)$ and $C_{\rm fs}^{[N]}(t)$ in the limits $t,N\to\infty$. There are many more of these spikes in multi-dimensions than in one-dimension, since there is a spike for every possible pair of (open) reactant or product channels.  However, it is possible to isolate each off-diagonal spike by constructing angular functions $F$ (see Sec.~V and Appendix D) in the space orthogonal to ${\overline f}({\ppi})$. It then follows that each off-diagonal spike in $A_N({\ppi})$ contributes zero to $C_{\rm fs}^{[N]}(t)$ in the limits $t,N\to\infty$, if there is no recrossing of surfaces orthogonal to $f({\bf q})$ in the space $\ssi$. 

Hence we have obtained the same result in multi-dimensions as in one-dimension: that the RPMD-TST rate is equal to the exact quantum rate if there is no recrossing of the dividing surface, nor of any surface orthogonal to it in  an $(N\!-\!1)$-dimensional space orthogonal to $f({\bf q})$, which describes (in the \shortt limit) the
 fluctuations  in the polymer-bead positions along the reaction coordinate $\sigma(q_1,\dots,q_F)$. It is impossible to recross these
  orthogonal surfaces in the classical (i.e.\ high-temperature limit), where RPMD-TST thus reduces to classical TST.

\section{Conclusions}
\label{sec:con}

We have shown  that quantum TST (i.e.\ RPMD-TST)  is related to the exact quantum rate in the same way that classical TST is related
to the exact classical rate; i.e.\
quantum TST is exact in the absence of recrossing. 
Recrossing in quantum TST is more complex than in classical TST, since, in addition to recrossing of the ring-polymer dividing surface, one must also consider recrossing through surfaces that describe fluctuations in
the positions of the polymer beads along the reaction coordinate. Such additional recrossing disappears in the classical and parabolic barrier limits, and thus becomes important only
 at temperatures below the cross-over to deep tunnelling. Previous RPMD-TST calculations\cite{jor} indicate that the resulting loss in accuracy increases slowly as the temperature is reduced below cross-over, such that quantum TST remains within a factor of two of the exact rate at temperatures down to below half the cross-over temperature. However, it is clear that further work will be needed in order to predict quantitatively how far one can decrease the temperature below cross-over before quantum TST breaks down (which will always happen
 at a sufficiently low temperature). 
 
Just as with classical TST, quantum TST will not work for indirect reactions, such as those involving long-lived intermediates, or diffusive dynamics (e.g.\ the high-friction regimes of the quantum Kramers problem\cite{hang}).  However, this leaves a vast range of chemical reactions for which quantum TST is applicable, and for which it will give an excellent approximation to the exact quantum rate. The findings in Part I and in this article thus validate the already extensive (and growing) body of results from RPMD rate-simulations\cite{rates,refined,azzouz,bimolec,ch4,mustuff,anrev,yury,tommy1,tommy2,tommy3,stecher,guo} (which give a lower bound to the RPMD-TST rate), as well as results obtained using the older centroid-TST method\cite{gillan1,gillan2,centroid1,centroid2,schwieters,ides} (which is a special case of RPMD-TST\cite{jor,cent}).

\begin{acknowledgments}
TJHH is supported by a Project Studentship from the UK Engineering and
Physical Sciences Research Council. 
\end{acknowledgments}

\section*{APPENDIX A: Long-time limit of the hybrid flux-side time-correlation function} 
\label{app:a}
\renewcommand{\theequation}{A\arabic{equation}}
\setcounter{equation}{0}

Here we derive \eqn{thicky}, which states  that ${\overline C}_{\rm fs}^{[N]}(t)$ gives the exact quantum rate in the \longt limit. We use the property that 
\begin{align}
{\overline C}_{\rm fs}^{[N]}(t)= {\overline C}_{\rm sf}^{[N]}(t)=-{d\over d t}{\overline C}_{\rm ss}^{[N]}(t) 
\end{align}
where ${\overline C}_{\rm sf}^{[N]}(t)$ and ${\overline C}_{\rm ss}^{[N]}(t)$  are the side-flux and side-side time-correlation functions corresponding to ${\overline C}_{\rm fs}^{[N]}(t)$. We then write ${\overline C}_{\rm sf}^{[N]}(t)$ as
\begin{align}
{\overline C}_{\rm sf}^{[N]}(t) = & \int\! d{\bf q}\, \int_{-\infty}^\infty\! d\Delta_1\,h[f({\bf q})]\nonumber\\ 
\times&\expect{q_{1}-\Delta_{1}/2|e^{-\beta_N{\hat H}}|q_2}\nonumber\\
\times&\prod_{i=3}^{N}\expect{q_{i-1}|e^{-\beta_N{\hat H}}|q_i}\nonumber\\
\times&\expect{q_N|e^{-\beta_N{\hat H}}|q_1+\Delta_1/2}
\nonumber\\ 
\quad\times &\expect{q_1+\Delta_1/2|e^{i{\hat H}t/\hbar}{\hat F}(q^\ddag) e^{-i{\hat H}t/\hbar}|q_1-\Delta_1/2}
\label{eq:cssN} 
\end{align}
where ${\hat F}(q^\ddag)$  is the flux operator\cite{MST}
\begin{align}
{\hat F}(q^\ddag) = {1\over 2 m}\left[\hat p\delta(q-q^\ddag) + \delta(q-q^\ddag)\hat p \right]
\end{align}
and insert identities of the form $e^{i{\hat H}t/\hbar}e^{-i{\hat H}t/\hbar}$ to obtain
\begin{align}
{\overline C}_{\rm sf}^{[N]}(t) = &\int\! d{\bf q}\,\int_{-\infty}^\infty\! d\Delta_1\,h[f({\bf q})]\nonumber\\ 
\times&\expect{q_{1}-\Delta_{1}/2|e^{i{\hat H}t/\hbar}e^{-\beta_N{\hat H}}e^{-i{\hat H}t/\hbar}|q_2}\nonumber\\
\times&\prod_{i=3}^{N}\expect{q_{i-1}|e^{i{\hat H}t/\hbar}e^{-\beta_N{\hat H}}e^{-i{\hat H}t/\hbar}|q_i}\nonumber\\
\times&\expect{q_N|e^{i{\hat H}t/\hbar}e^{-\beta_N{\hat H}}e^{-i{\hat H}t/\hbar}|q_1+\Delta_1/2}
\nonumber\\ 
\quad\times &\expect{q_1+\Delta_1/2|e^{i{\hat H}t/\hbar}{\hat F}(q^\ddag) e^{-i{\hat H}t/\hbar}|q_1-\Delta_1/2}
\label{bob} 
\end{align}
This allows us to take  the \longt limit of ${\overline C}_{\rm sf}^{[N]}(t)$  by using \eqn{dragons} together with the relation
\begin{align}
\hat\Omega_+\ket{p_i}= \ket{\phi^+_{p_i}}
\end{align}
where
\begin{align}
\hat\Omega_+\equiv\lim_{t\to\infty}e^{-i\hat H t/\hbar}e^{i\hat K t/\hbar}
\end{align}
and $\ket{\phi^+_{p_i}}$ is a (reactive) scattering wave function with incoming boundary conditions.\cite{taylor} We then
obtain
\begin{align}
{\overline C}_{\rm sf}^{[N]}(t) = \int\! d{\bf p}\, &\int_{-\infty}^\infty\! dp'_1\,h\!\left[{\overline f}\left({p_1+p_1'\over 2},p_2,\dots,p_N\right)\right]\nonumber\\ 
\times&\expect{\phi^+_{p_1'}|e^{-\beta_N{\hat H}}|\phi^+_{p_{2}}}\nonumber\\
\times&\prod_{i=3}^{N}\expect{\phi^+_{p_{i-1}}|e^{-\beta_N{\hat H}}|\phi^+_{p_i}}\nonumber\\
\times&\expect{\phi^+_{p_{N}}|e^{-\beta_N{\hat H}}|\phi^+_{p_1}}\expect{\phi^+_{p_1}|{\hat F}(q^\ddag) |\phi^+_{p_1'}}
\label{jupiter} 
\end{align}
From the orthogonality of the scattering eigenstates,\cite{taylor} we obtain
\begin{align}
\expect{\phi^+_p|e^{-\beta_N{\hat H}}|\phi^+_{p'}}=e^{-p^2\beta_N/2m}\delta(p-p')
\end{align}
We also know that
\begin{align}
h\left[{\overline f}(p,p,\dots,p)\right]=h(p)
\end{align}
(since otherwise $f({\bf q})$ would not correctly distinguish between reactants and products in the limit \longta).
 We thus obtain
\begin{align}
\lim_{t\to\infty}{\overline C}_{\rm sf}^{[N]}(t) &= \int_{-\infty}^{\infty}\! dp\, e^{-p^2\beta/2m} h(p)
\expect{\!\phi^+_p|{\hat F}(q^\ddag) |\phi^+_p\!}
\label{pip} 
\end{align}
which is the \longt limit of the Miller-Schwarz-Tromp flux-side time-correlation function, \cite{MST} from which we obtain \eqn{thicky}.

\section*{APPENDIX B: Derivation of the structure of $A_N({\bf p})$} 
\label{app:b}
\renewcommand{\theequation}{B\arabic{equation}}
\setcounter{equation}{0}

Here we derive \eqn{piggy} of Sec.~IV. 
We first define a special type of side-side time-correlation function,
\begin{align}
{ P}_{l}^{[N]}({\bf E},t)=\int\! & d{\bf q}\, \int\! d{\bf z}\,\int\! d{\bf \Delta}\,\no\\
\times & h[f({\bf q})]
\left[\prod_{i=1,i\ne l}^N
h(z_i-q^\ddag)\right]\nonumber\\ 
\times&\prod_{i=1}^{N}\expect{q_{i-1}-\Delta_{i-1}/2|e^{-\beta_N{\hat H}}|q_i+\Delta_i/2}
\nonumber\\ 
\quad\times &\expect{q_i+\Delta_i/2|e^{i{\hat H}t/\hbar}\delta({\hat H}-E_i)|z_i}\no\\
\times&\expect{z_i|e^{-i{\hat H}t/\hbar}|q_i-\Delta_i/2}
\label{tebbit} 
\end{align}
where ${\bf E}\equiv\{E_1,E_2,\dots,E_N\}$, and then consider its \longt time-derivative,
\begin{align}
{Q}_{l}^{[N]}({\bf E})=\lim_{t\rightarrow\infty}{d\over dt}{\overline P}_{l}^{[N]}({\bf E},t)
\end{align}
We can obtain two equivalent expressions for ${Q}_{l}^{[N]}({\bf E})$, by evaluating it as either a flux-side or a side-flux time-correlation function. 
The flux-side version is
\begin{align}
{Q}_{l}^{[N]}({\bf E})=\int\! d{\bf q}\, &\int\! d{\bf p}\,\int\! d{\bf \Delta}\,{\cal \hat F}[f({\bf q})]
\left[\prod_{i=1,i\ne l}^N
h(p_i)\right]\nonumber\\ 
\times&\prod_{i=1}^{N}\expect{q_{i-1}-\Delta_{i-1}/2|e^{-\beta_N{\hat H}}|q_i+\Delta_i/2}
\nonumber\\ 
\quad\times &\expect{q_i+\Delta_i/2|\delta({\hat H}-E_i)|\phi^-_{p_i}}\no\\
\times&\expect{\phi^-_{p_i}|q_i-\Delta_i/2}
\label{rip_thatcher} 
\end{align}
which gives
 \begin{align}
  |p_l|^{-1} & A_N(p_1,\dots,p_l,\dots,p_N)\no \\
  & + |{\w p}_l|^{-1}A_N(p_1,\dots,{\w p}_l,\dots,p_N)\nonumber\\
  &\qquad ={{Q}_{l}^{[N]}({\bf E})\over m^N}\prod_{i=1,i\ne l}^N |p_i|\ \ \ \ \text{\rm if ${\w p}_l$ real} \label{dennis}
     \end{align}
and  
\begin{align}
  |p_l|^{-1}&A_N(p_1,\dots,p_l,\dots,p_N)= \no \\
&{{Q}_{l}^{[N]}({\bf E})\over m^N}\prod_{i=1,i\ne l}^N |p_i|  \ \ \ \text{\rm if ${\w p}_l$ imaginary}
\label{mennis}
   \end{align}
The side-flux version is
\begin{align}
{ Q}_{l}^{[N]}({\bf E})=\int\! d{\bf s}\, &\int\! d{\bf s}'\,h[f({\bf s+s'})]\no\\
\times&\left[\prod_{i=1}^{N}\expect{\phi^+_{s_{i-1}'}|e^{-\beta_N{\hat H}}|\phi^+_{s_i}}\right]\no\\
\times&\expect{\phi^+_{s_l}|\delta({\hat H}-E_l)|\phi^+_{s_l'}}\no\\
\times &\sum_{j=1,j\ne l}^N\ \expect{\phi^+_{s_j}|\delta({\hat H}-E_j){\hat F}(q^\ddag)|\phi^+_{s_j'}}\no\\ 
\quad\times &\prod_{\substack{i=1,i\ne l\\i\ne j}}^N
\expect{\phi^+_{s_i}|\delta({\hat H}-E_i){\hat h}(q^\ddag)|\phi^+_{s_i'}}
\label{maggie_morta} 
\end{align}
The second to fourth lines in this expression contain the $\delta$-functions,
\begin{align}
\delta(s_l-s'_l)\prod_{i=1}^{N}\delta(s'_{i-1}-s_i)\delta[E^+(s_i)-E_i]
\end{align}
where $E^+(s_i)$ is defined the other way round to $E^-(p_i)$ of \eqn{nofood}, and where the $\delta$-functions in $s_i $ and $s'_i$ result from the orthogonality of the scattering states $\ket{\phi^+_{s}}$.\cite{taylor} Integrating over $s_i $ and $s'_i$, we obtain
\begin{align}
{ Q}_{l}^{[N]}({\bf E})=b_N({\bf p})\delta(E_{l+1}-E_l)
\end{align}
where $b_N({\bf p})$ is some function of ${\bf p}$ (which we do not need to know explicitly). Substituting this expression into \eqnn{dennis}{mennis}, we obtain
 \begin{align}
  |p_l|^{-1} & A_N(p_1,\dots,p_l,\dots,p_N)\no\\
  &+ |{\w p}_l|^{-1}A_N(p_1,\dots,{\w p}_l,\dots,p_N)\nonumber\\
  &\qquad =c_N({\bf p})\delta(E_{l+1}-E_l)
\ \ \ \ \text{\rm if ${\w p}_l$ real}
\label{brutus}
 \end{align}
 and
 \begin{align}
  |p_l|^{-1}&A_N(p_1,\dots,p_l,\dots,p_N)\no\\
  & =c_N({\bf p})\delta(E_{l+1}-E_l)
\ \ \ \ \text{\rm if ${\w p}_l$ imaginary}
\label{crutus}
 \end{align}
where $c_N({\bf p})$ is some function of ${\bf p}$. This derivation was obtained for the case that $p_i>0$, $i\ne l$, but can clearly be repeated for all combinations of positive and negative $p_i$  [by replacing various $h(z_i-q^\ddag)$
  by $h(-z_i+q^\ddag)$]. Hence \eqnn{brutus}{crutus} holds for all ${\bf p}$. 

Now, we can obtain alternative expressions for the righthand side of \eqnn{brutus}{crutus} by adding and subtracting sequences of terms that correspond to following different paths through the hypercube. Consider, for example (for the case that ${\w p}_i, {\w p}_j$ are both real), the sequence 
\begin{align}
  |p_j|^{-1}&A_N(p_1,\dots,p_i,\dots,p_j,\dots,p_N)\no\\
   &+ |{\w p}_j|^{-1}A_N(p_1,\dots,{p}_i,\dots,{\w p}_j,\dots,p_N)\nonumber\\
   &\qquad =X_N({\bf p})\delta(E_{j+1}-E_j)\no\\
  |p_i|^{-1}&A_N(p_1,\dots,p_i,\dots,{\w p_j},\dots,p_N)\no\\
   &+ |{\w p}_i|^{-1}A_N(p_1,\dots,{\w p}_i,\dots,{\w p}_j,\dots,p_N)\nonumber\\
   &\qquad =Y_N({\bf p})\delta(E_{i+1}-E_i)\no\\
  |{\w p}_j|^{-1}&A_N(p_1,\dots,{\w p}_i,\dots,{\w p}_j,\dots,p_N)\no\\
  &+ |{p}_j|^{-1}A_N(p_1,\dots,{\w p}_i,\dots,{p}_j,\dots,p_N)\nonumber\\
   &\qquad =Z_N({\bf p})\delta(E_{j+1}-E_j)
\label{bag-lady}
\end{align}
where each of $X_N({\bf p}),Y_N({\bf p}),Z_N({\bf p})$ is some (different) function of ${\bf p}$. Combining these expressions, we obtain
\begin{align}
  |p_i|^{-1}&A_N(p_1,\dots,p_i,\dots,p_N)\no\\
  & + |{\w p}_i|^{-1}A_N(p_1,\dots,{\w p}_i,\dots,p_N)\nonumber\\
  &\qquad=P_N({\bf p})\delta(E_{i+1}-E_i)+Q_N({\bf p})\delta(E_{j+1}-E_j)
\label{thugus}
 \end{align}
where $P_N({\bf p})=-|p_j||{\w p}_j|^{-1}Y_N({\bf p})$,
 and $Q_N({\bf p})=|p_j|\left[|p_i|^{-1}X_N({\bf p})+|{\w p}_i|^{-1}Z_N({\bf p})\right]$. \cite{qzero} We can repeat this procedure for each of the $N-1$ different values of $j\ne i$. Because the resulting set of coefficients $P_N({\bf p})$ and $Q_N({\bf p})$ are linearly independent\footnote{It is conceivable that these coefficients might become linearly dependent at some value of ${\bf p}$ but these would be isolated points and thereby contribute nothing to the integral.}, it follows that
 \begin{align}
   |p_i|^{-1}&A_N(p_1,\dots,p_i,\dots,p_N)\no\\
   &+ |{\w p}_i|^{-1}A_N(p_1,\dots,{\w p}_i,\dots,p_N)\nonumber\\
   &\qquad=d_N({\bf p})\prod_{i=1}^{N-1}\delta(E_{i+1}-E_i) \ \ \ \ \text{\rm if ${\w p}_i$ real}
   \label{belgrano}
 \end{align}
 and 
  \begin{align}
   |p_i|^{-1}&A_N(p_1,\dots,p_i,\dots,p_N)\no\\
   &=d_N({\bf p})\prod_{i=1}^{N-1}\delta(E_{i+1}-E_i) \ \ \ \ \text{\rm if ${\w p}_i$ imaginary}
   \label{borax}
 \end{align}
where $d_N({\bf p})$ is some function of ${\bf p}$. From this, we obtain \eqn{piggy} of Sec.~IV.

\section*{APPENDIX C: Cancellation of the term $r_N({\bf p})$  in the limit \ntya} 
\label{app:c}
\renewcommand{\theequation}{C\arabic{equation}}
\setcounter{equation}{0}

Because the function ${\overline{f}({\bf p})}$ must vary smoothly with imaginary time and converge in the limit \ntya, it can be rewritten as a function of a finite number $K$ of the linear combinations
\begin{align}
{\ov P}_i=\sum_{j=1}^N T_{ij}p_j\ \ \ \ \ \ i=1,\dots,K
\end{align}
in which $T_{ij}\sim N^{-1}$ (i.e.~${\ov P}_i$ is normalised such that it converges in the limit $N\to\infty$; e.g.~$T_{0j}= N^{-1}$ corresponds to the centroid).
 It then follows that ${\partial f({\bf p})/ \partial p_j}\sim N^{-1}$, and hence that
\begin{align}
\lim_{N\rightarrow\infty}{\overline{f}}(p_1,\dots,{\w p}_j,\dots,p_N)={\overline{f}}({\bf p})
+({\w p}_j-p_j)
{\partial f({\bf p})\over \partial p_j}
\end{align}
 provided the range of ${\w p}_j-p_j$ is finite [which it is because $r_N({\bf p})$ contains Boltzmann factors]. 
 We can therefore write the \nty limit of \eqn{residue} as
 \begin{align}
\lim_{N\rightarrow\infty}&S(N)=\no\\
&\int_{-\infty}^\infty\! dp_1\dots\int_{-\infty}^\infty\! dp_{j-1}\int_{0}^\infty\! dp_{j}\dots\int_{0}^\infty\! dp_{N}\,r_N({\bf p})\no\\
&\qquad \times ({\w p}_j-p_j)
{\partial f({\bf p})\over \partial p_j}\delta[{\overline{f}}(p_1,\dots,p_j,\dots,p_N)]
\end{align}
which shows that the volume sandwiched between the two heaviside functions becomes a strip of width
 $({\w p}_j-p_j){\partial f({\bf p})/ \partial p_j}\sim N^{-1}$ in the limit \ntya.  
 
\section*{APPENDIX D: Isolating the off-diagonal spikes for $N>3$} 
\label{app:d}
\renewcommand{\theequation}{D\arabic{equation}} 
\setcounter{equation}{0}

It is straightforward to generalize the result obtained for $N=3$ and $f({\bf q})={\ov q}_0$ in Sec.~V.B  to general $N$ and to any (cyclically invariant) choice of $f({\bf q})$. 

We consider first the special case of a centroid dividing surface [$f({\bf q})={\ov q}_0$]. The space orthogonal to ${\ov q}_0$
 can be represented by orthogonal linear combinations $Q_i$, $i=1,\dots N-1$ of $q_i$, analogous to $Q_x$ and $Q_y$ in Sec.~V.B. We can then define a generalized radial dividing surface
\begin{equation}
g_r({\bf q})=\sqrt{\sum_{i=1}^{N-1}Q_i^2}-r^\ddag
\end{equation}
(which is invariant under cyclic permutation of the $q_i$) and generalized angular dividing surfaces
\begin{equation}
g_F({\bf q})=F\!\left[\phi({Q_X,Q_Y}) \right]
\end{equation}
with
\begin{align}
\phi(Q_X,Q_Y) &= \arctan(Q_Y/Q_X)\label{phiphi}
\end{align} 
where $(Q_X,Q_Y)$ can be chosen to be any mutually orthogonal  pair of linear combinations of the $Q_i$.
From \eqn{baffi}, the \longt limits of $g_r({\bf q})$ and $g_F({\bf q})$ are
\begin{equation}
{\ov g}_r({\bf p})=\lim_{\epsilon\to 0}\sqrt{\sum_{i=1}^{N-1}P_i^2}-\epsilon
\end{equation}
and
\begin{equation}
g_F({\bf p})=F\!\left[\phi(P_X,P_Y) \right]
\end{equation}
where $P_i$ and $(P_X,P_Y)$ are the linear combinations of $p_i$ analogous to $Q_i$ and $(Q_X,Q_Y)$.
We can then proceed as for the $N=3$ example in Sec.~V.B. Substituting ${\ov g}_r({\bf p})$ into \eqn{lunghi}, we obtain the constraint that the spikes along the centroid axis contribute zero (since ${\ov g}_r({\bf p})$ encloses these spikes only). This leaves us free to construct angular dividing surfaces $g_F({\bf q})$  in various planes $(Q_X,Q_Y)$ (which need not be mutually orthogonal) in order to enclose individual off-diagonal spikes. \cite{exist} \Eqn{lunghi} then gives a set of constraints, each stating that the contribution to $A_N({\bf p})$ from one of these spikes is zero if there is no recrossing of any surface orthogonal to~$f({\bf q})$.

This reasoning can be applied with slight modification to  a general (cyclically invariant) dividing surface $f({\bf q})$. By construction, such  a surface reduces to a function of just the centroid near the centroid axis, and hence there exists a radial coordinate in the $(N\!-\!1)$-dimensional curvilinear space orthogonal to $f({\bf q})$ which reduces to $g_r({\bf q})$ close to the centroid axis. We therefore obtain the constraint that the spikes along the centroid axis contribute zero, and are then free to isolate each of the off-diagonal spikes by using curvilinear generalisations of the angles $\phi$, which sweep over curvilinear surfaces that are
orthogonal to $f({\bf q})$, and which reduce to the form of \eqn{phiphi} close to the centroid axis.

%
%
%
%
%
%
%
%
%
%
%
%
%

\begin{thebibliography}{10}
\bibitem{vlow} A centroid dividing surface will break down also for a symmetric barrier at very low temperatures, when it becomes necessary to include other (even) normal modes of the ring-polymer in the dividing surface; see ref.\ \onlinecite{jor}.
\bibitem{23} This supplies the necessary proof of Eq.~(23) of Part I, which was deferred to the present article.
\bibitem{bc} Meaning that $\expect{x|\phi^-_p}$ has the form
$\lim_{x\to\infty}\expect{x|\phi^-_{p}}=\expect{x|p}+R(p)\expect{x|-\!p}$ for $p>0$, where $R(p)$ is the reflection coefficient; see 
ref.\ \onlinecite{taylor}.
\bibitem{centy} 
One can show that $r_N({\bf p})$ cancels exactly for finite, even, $N$, in the special case of a centroid dividing surface. This follows from the property that ${\ov p}_0({\bf p})=-{\ov p}_0(-{\bf p})$.
\bibitem{unless} This statement would be invalid if the integral of $r_N({\bf p})$ over a subcube were to grow at least as rapidly as $N^N$. One can show that this is not the case, by writing out  $A_N({\bf p})$  in its side-flux form.
\bibitem{cancel} We are ignoring that $r_N({\bf p})$ does not cancel out perfectly for $N=3$, since our aim is to illustrate 
 the use of the constraints on the spikes, which are then applied in the \nty limit (where $r_N({\bf p})$ does cancel out).
\bibitem{symmetric} For the case of a symmetric barrier, the spikes along the centroid axis will sum to zero even if there is recrossing, on account of \eqn{goody}.
\end{thebibliography}
\end{document}